\documentclass[aps,prl,twocolumn,superscriptaddress]{revtex4}

\usepackage{amsfonts}%
\usepackage{amsmath}%
\usepackage{amssymb}%
\usepackage{graphicx}

\begin{document}

\title{Strong constraints on models that explain the violation of Bell inequalities with hidden superluminal influences}
\author{Valerio Scarani}
\affiliation{Centre for Quantum Technologies, National University of Singapore, 3 Science Drive 2, Singapore 117543}
\affiliation{Department of Physics, National University of Singapore, 2 Science Drive 3, Singapore 117542}
\author{Jean-Daniel Bancal}
\affiliation{Centre for Quantum Technologies, National University of Singapore, 3 Science Drive 2, Singapore 117543}
\author{Antoine Suarez}
\affiliation{Center for Quantum Philosophy, Berninastrasse 85, 8057 Zurich/Switzerland}
\author{Nicolas Gisin}
\affiliation{Group of Applied Physics, University of Geneva, Switzerland}
\keywords{}
\pacs{PACS number}

\begin{abstract}
We discuss models that attempt to provide an explanation for the violation of Bell inequalities at a distance in terms of hidden influences. These models reproduce the quantum correlations in most situations, but are restricted to produce local correlations in some configurations. The argument presented in [Bancal et al. Nature Physics 8, 867 (2012)] applies to all of these models, which can thus be proved to allow for faster-than-light communication. In other words, the signalling character of these models cannot remain hidden.
\end{abstract}
\maketitle

Since the advent of Bell inequalities, quantum predictions are known to be incompatible with locally causal theories~\cite{nouvellecuisine}. Experimental evidence of Bell inequality violation, in agreement with quantum predictions, have motivated the search for alternative ways to explain these violations. One such explanation invokes faster than light influences propagating in a preferred reference frame~\cite{Eberhardt89}. It accounts for every Bell inequality violation that can be observed experimentally, while restoring a notion of information propagation in space-time. However, this model was recently shown to lead to faster-than-light communication in some spacetime configuration~\cite{Bancal12}; see also~\cite{Gisin12} for an elaboration on the importance of this result by one of us.

The argument used in~\cite{Bancal12} to reach this conclusion relies on a special kind of Bell inequalities, termed hidden influence inequalities. Interestingly, these inequalities allow not only to rule out finite-speed influence models, but can also be used to exclude similar alternative models. Here, we discuss this extension for two models:
\begin{itemize}
\item The finite-distance model, which allows for simultaneous events to influence themselves and thus be arbitrarily correlated if they are close enough to each other, but predicts that quantum correlations disappear after a given distance in a preferred frame.
\item The multisimultaneity model, which considers the rest frame of the measurement stations as the relevant frames for the signalling~\cite{SS97}.
\end{itemize}

The models considered here attempt to describe observable nonlocal correlations by a signalling mechanism defined in some suitable frame. Each model further specifies this mechanism, and ends up departing from the predictions of quantum theory in some configurations. Aside from the specifics that we shall describe in detail below, these tentative models thus share the following properties:
\begin{itemize}
\item[(i)] They assume that the predictions of quantum theory are correct in most spacetime configurations.
\item[(ii)] In some configurations, however, the model predicts a disappearance of nonlocal correlations whereas quantum theory predicts a violation of some Bell inequality. These models do not consider the possibility that supra-quantum no-signaling correlations will ever be observed.
\end{itemize}

The aim of this paper is to show that the influences in such models appear in their experimental predictions and allow for faster-than-light communication between users, in flagrant contradition with relativity.

Note that the models considered here do not fall under the claim of Colbeck and Renner that any no-signalling model reproducing the predictions of quantum theory is effectively equivalent to the latter~\cite{Colbeck11}. These authors work under the assumption that all the resources obey the nosignalling principle, whereas models with influences clearly violate this principle. Additionally, they assume that the quantum correlations are recovered in every configuration, which is not the case here.


\ \\
\emph{No-signalling and hidden influence inequalities. }
Let us recall the link between the Bell inequality we consider here and the no-signalling condition. In a multipartite Bell-type experiment in which no information can be transmitted from one party to another one by its choice of measurement settings, the observed statistics satisfy the so-called no-signalling conditions, i.e.:
\begin{equation}\label{eq:nosig}
\sum_a P(ab\ldots|xy\ldots)  = P(b\ldots|y\ldots)\ \forall\ x,
\end{equation}
and similarly for sums on the other outcomes. Here we denote the parties' measurement settings by $x,y,\ldots$, their respective outcomes by $a,b,\ldots$, and $P$ stands for probability. The no-signalling conditions \eqref{eq:nosig} hold true for the result of any quantum experiment.

The correlations produced by an influence model are local in configurations in which no influences can be exchanged. 
Specifically, let us consider a four-partite configuration in which two parties, say Alice and Dave, measure in advance, one after the other one, but the other two, Bob and Charlie, perform their measurements in a situation where a departure from the quantum case is predicted for the considered model. Then the following locality condition must be satisfied by the bipartite correlations of Bob and Charlie conditionned on Alice and Dave:
\begin{equation}\label{eq:locality}
P(bc|yz,adxw)=\sum_\lambda q(\lambda)P(b|y,adxw,\lambda)P(c|z,adxw,\lambda).
\end{equation}

The hidden influence inequality $S\leq 7$ described in~\cite{Bancal12} captures a combination of the two conditions above: it is satisfied whenever both \eqref{eq:nosig} and \eqref{eq:locality} hold. For a model granting that \eqref{eq:locality} holds in a given situation, because of (ii), violation of this inequality thus implies that the nosignalling condition is violated.

In order to demonstrate that a model of the kind considered here leads to faster-than-light communication, it is thus sufficient to check that the following two conditions are satisfied in a configuration in which the model predicts that condition (2) is satisfied (i.e. in which the nonlocality between Bob and Charlie is lost):
\begin{itemize}
\item[(a)] the considered model predicts a violation of the inequality $S\leq 7$
\item[(b)] the signalling in the correlations $P(abcd|xyzw)$ can be used to communicate faster than light.
\end{itemize}
To check these conditions, it is useful to take advantage of the fact that the expression $S$ only involves correlations between Alice-Bob-Dave and Alice-Charlie-Dave. One then only needs to check, that these two marginals are correctly reproduced by the model in order to fulfill (a) since quantum theory allows for a violation of this inequality. Also, if the model indeed predicts these tripartite marginals to take the same value than the quantum ones, then the signalling in the correlations $P(abcd|xyzw)$ must show up in one of the two remaining tripartite marginals, i.e. in the Alice-Bob-Charlie or the Bob-Charlie-Dave one. Condition (a) and (b) can then be re-expressed as:
\begin{itemize}
\item[(a')] The Alice-Bob-Dave and Alice-Charlie-Dave marginals produces by the model in the considered configuration are the same as the ones predicted by quantum theory.
\item[(b')] There exist a point $A'$ that lies in the future light-cone of Alice, Bob and Charlie, but outside of Dave's future lightcone; and similarly for a point $D'$.
\end{itemize}

Before introducing the models we want to consider here and checking the above conditions on them in an appropriate space-time configuration, let us recall in the next paragraph how these steps are performed in the case of the $v$-causal models.

\ \\
\emph{Finite speed influences in a universal preferred frame. }

\begin{figure}
\includegraphics[width=0.7\columnwidth]{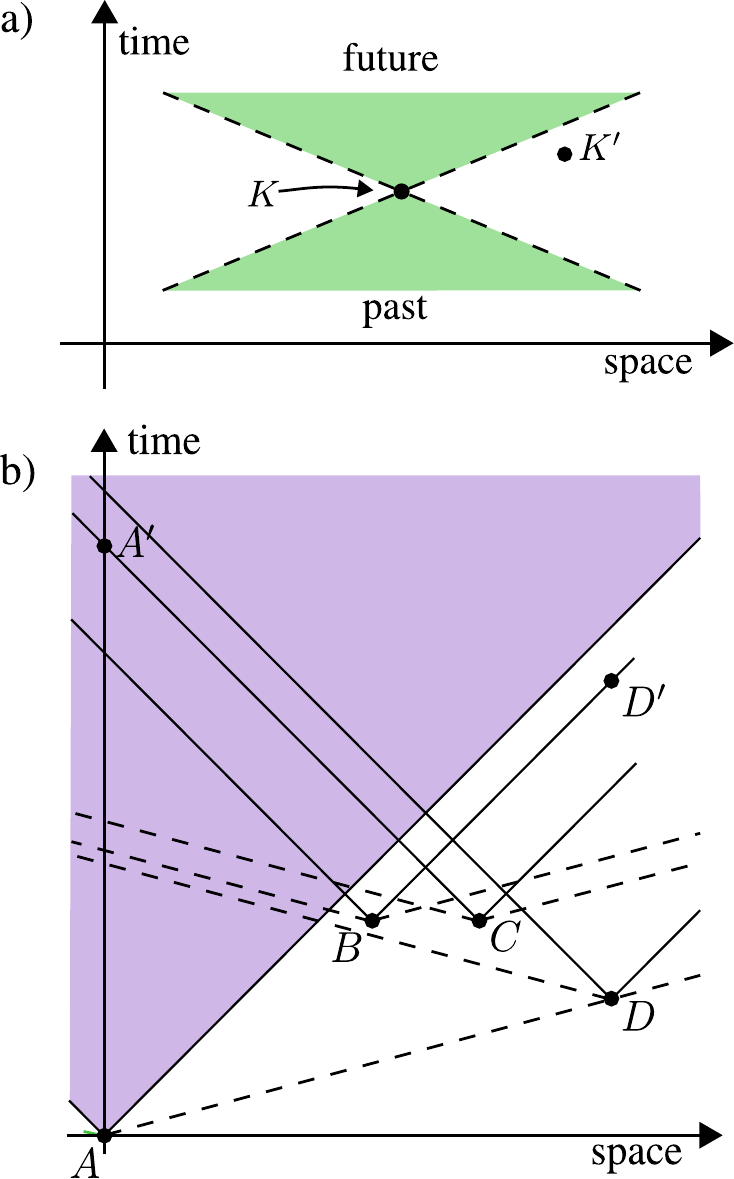}
\caption{Figure in the preferred reference frame. a) The colored region bounded by the dashed lines represents the past and future $v$-cone of event $K$. All events that are in this region can send/receive influences to/from $K$ and can thus share nonlocal correlations with $K$. However, only local correlations can arise between points $K$ and $K'$ which lie outside each other's $v$-cones. b) space-time configuration which allows to conclude that $v$-causal models lead to faster-than-light communication. The full lines represent lightcones; the colored region is the future light-cone of $A$. (see~\cite{Bancal12} for details).}
\label{fig:finitespeed}
\end{figure}

$v$-causal models allow influences to propagate between events at the speed $v<\infty$, see Fig.~\ref{fig:finitespeed}a. They predict a departure from the quantum predictions when the measurements performed by several parties are sufficiently simultaneous to each other: only local correlations can be produced between systems measured close to simultaneously. This implies, in the configuration depicted in Fig.~\ref{fig:finitespeed}b, that the condition \eqref{eq:locality} must be satisfied. Nevertheless, one can check that the tripartite marginals Alice-Bob-Dave and Alice-Charles-Dave produced by the model in this configuration agree with the quantum predictions (point (a')), which implies that a violation of the hidden influence inequality can be produced by the model in this situation. Indeed, if Charlie decided to postpone his measurement for long enough, no measurements would be too simultaneous for the model and thus the Alice-Bob-Dave marginal produced should coincides with the quantum prediction. But since none of Alice, Bob, or Dave can learn whether Charlie postpones his measurement or not at the moment they produces their outcomes, in the situation of Fig.~\ref{fig:finitespeed}b, the Alice-Bob-Dave marginal must be identical to the quantum one. The same reasoning applies to the Alice-Charles-Dave marginal if Bob can also postpone his measurement, ensuring that both marginals follow with the quantum predictions. Finally, one can check directly that events $A'$ and $D'$ satisfy condition (b') (see Fig.~\ref{fig:finitespeed}b). This concludes the argument (see~\cite{Bancal12} for more details).

\ \\
\emph{The finite distance model in a universal preferred frame.}

If $v$-causal models lead to faster-than-light communication for any speed of the hidden influences $v<\infty$, this is not the case if $v=\infty$ since quantum correlations can always be recovered in this case. However, letting $v=\infty$ amounts to allow for an instantaneous connection between arbitrarily distant points in space. Is this the last option in absence of faster-than-light communication? Here, we consider a model which attempts to offers a less extreme picture. Namely, we consider what happens if instantaneous connections are allowed between points separated in space, but only up to a finite distance. We then show that this also leads to faster-than-light communication.

\begin{figure}
\includegraphics[width=0.65\columnwidth]{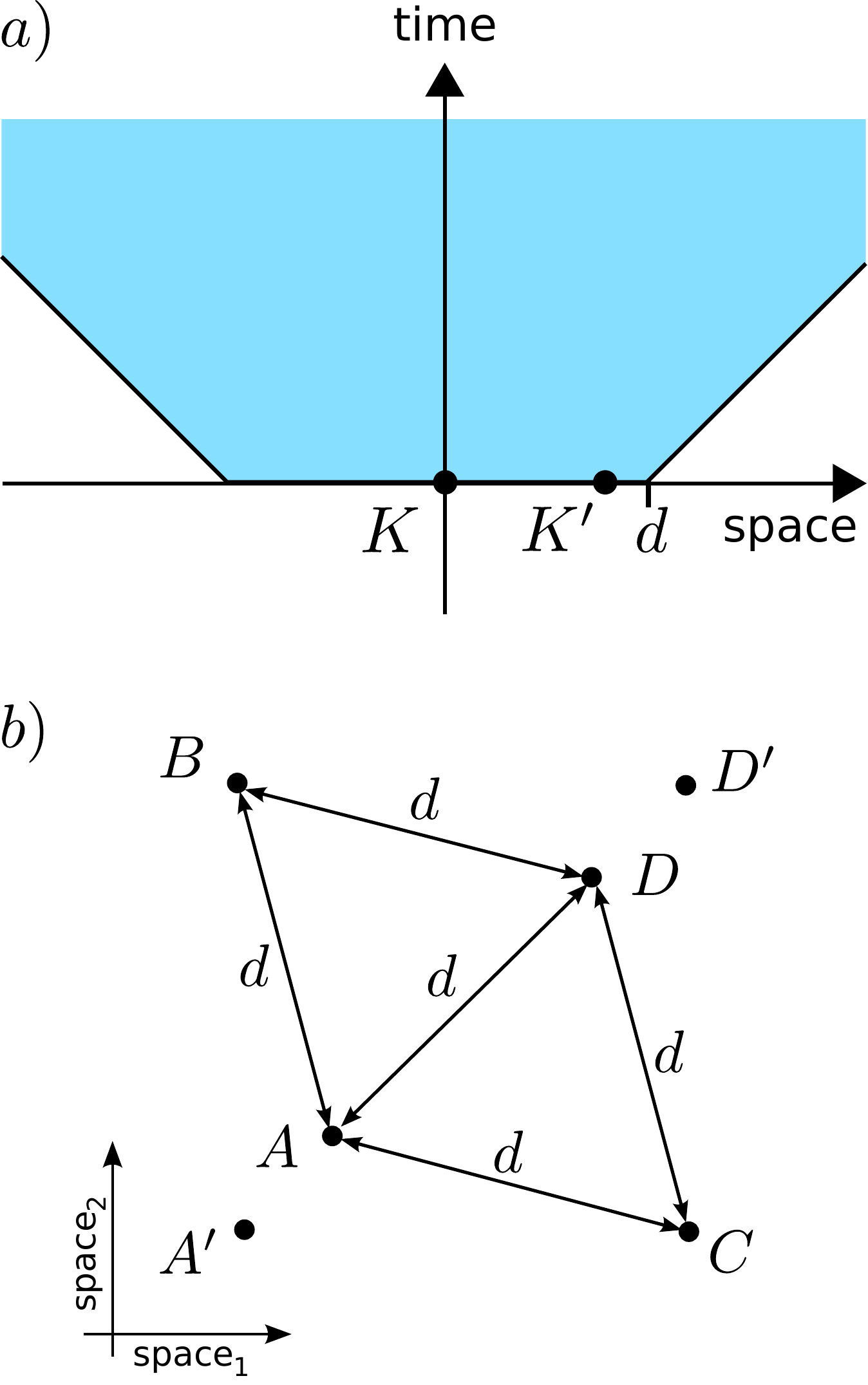}
\caption{Figure in the preferred reference frame. a) The colored region is the zone of influence of event $K$, i.e. the region of spacetime that can be influenced by $K$. Points $K$ and $K'$ can share arbitrary correlations because they are simultaneous in the preferred frame and separated by a distance smaller that $d$. b) Spatial configuration allowing to show that the finite-distance model leads to faster-than-light communication. Alice is separated from the other parties by a distance $d$, as well as Dave. Thus Bob and Charlie are separated by a distance $\sqrt{3}d>d$. Moreover, we let the parties measure at time $t_A=-2\epsilon$, $t_D =-\epsilon$ for $\epsilon>0$, and $t_B=t_C=0$. The points $A'$ and $D'$ are equidistant to $B$ and $C$ and at distance of at least $2\epsilon c(d-\epsilon c)/(d-4\epsilon c)$ from $A$ and $D$.}
\label{fig:finitedistance}
\end{figure}

For concreteness, we define our model in a preferred reference frame. In the finite distance model information about a measurement is instantaneously accessible to parties located in the neighboorhood of the measured system up to a distance $d$ (see Fig.~\ref{fig:finitedistance}a). As a result, such a model can produce arbitrary nosignalling correlations for any measurements performed by $n$ parties mutually separated two by two by a distance smaller than $d$ (independently of their individual measurement time). In particular, every Bell inequality violation observed so far can be explained in this way by choosing a large enough critical distance $d$.

In order to link the predictions of the finite-distance model with the ones predicted by quantum theory, we make the assumption that the model produces the expected quantum correlations at least in all situations where no two parties lie outside each other's zone of influence (see Fig.~\ref{fig:finitedistance}a).

The fact that the finite distance model leads to faster-than-light communication can be shown with the aid of the thought experiment depicted in Fig.~\ref{fig:finitedistance}b. In this configuration Alice and Dave measure before Bob and Charlie, and these last parties are further away than the critical distance $d$. The locality condition \eqref{eq:locality} is thus fulfilled by the model and one only needs to check that the points (a') and (b') are also satisfied in this configuration.

For the first point, one can see that the quantum Alice-Bob-Dave and Alice-Charlie-Dave marginals are correctly reproduced by the model with a similar argument as for $v$-causal models: if Bob (Charlie) can decide to postpone their measurements, then he can come in the zone of influence of the other party in such a way that the model produces the quantum correlations for all parties. So in particular the Alice-Charlie(Bob)-Dave marginal is the one predicted by quantum theory. 

For the second point, one can check directly that any information traveling at the speed of light from $A$ will reach the point $D'$  after the information coming at the same speed from $B$, $C$ and $D$, and similarly for $A'$. These two points thus satisfy condition (b'). It follows from our previous discussion that the model predicts the possibility of communicating faster than light in this configuration.


\ \\
\emph{The multisimultaneity model. }

\begin{figure}
\includegraphics[width=1\columnwidth]{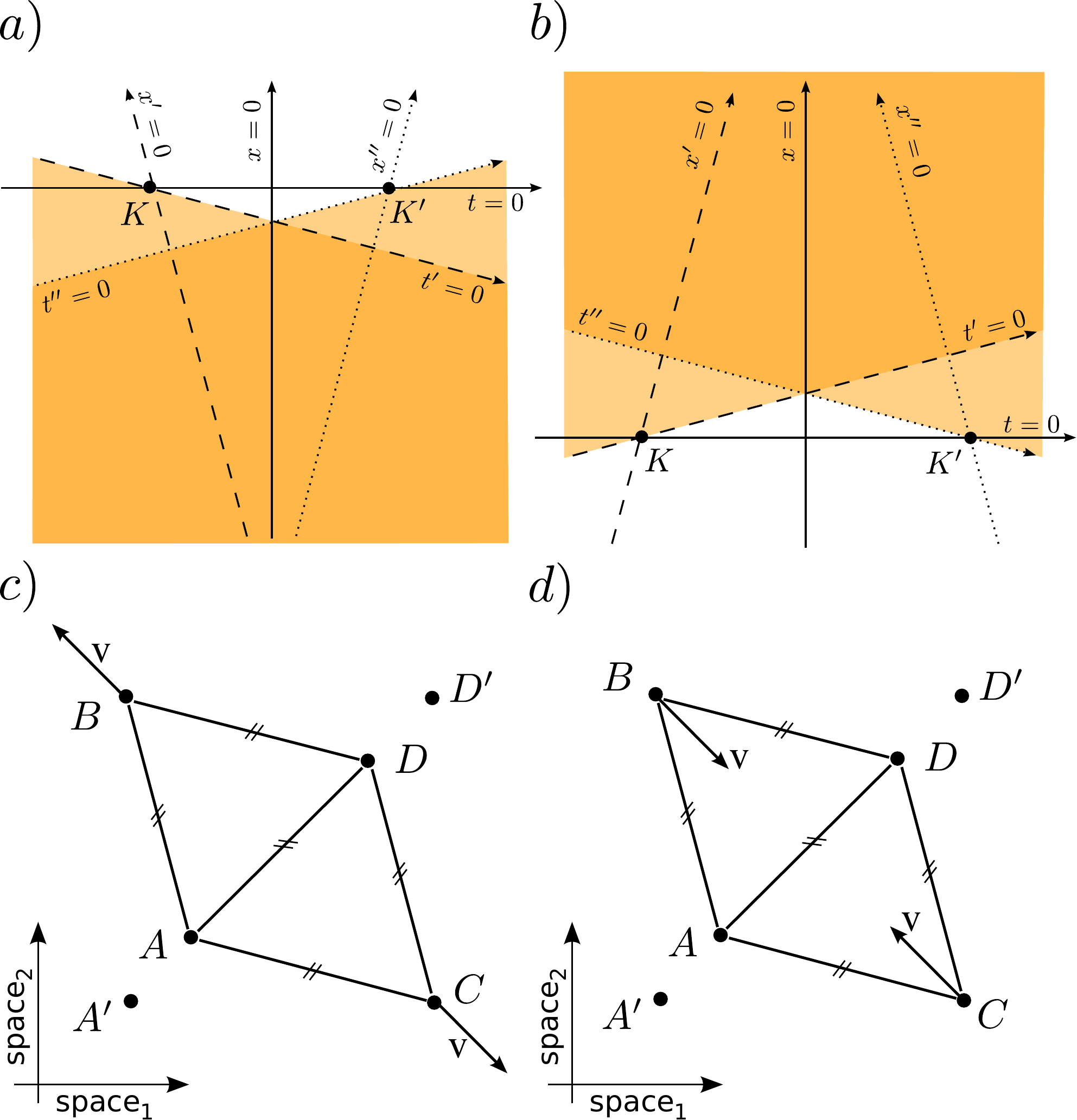}
\caption{a) and b) Before-before configuration for a two-particle multisimultaneity experiment of the first and second kind. The colored regions are the zones that can either influence events $K$ and $K'$, or be influenced by them. c) and d) Spatial configurations that allow one to show that these models are lead to faster-than-light communication. Points $A$, $B$, $C$, $D$ are positioned as in Fig.~\ref{fig:finitedistance}b. The measurement times are $t_A=t_D=-\sqrt{3}vd/(2c^2)$ and $t_B=t_C=0$ for Bob and Charlie moving at speed v$\ >0$. Points $A'$ and $D'$ are equidistant to $B$ and $C$ and at a distance larger than $\frac{dv(4\sqrt{3}c-3v)}{4c(c-\sqrt{3}v)}$ from $A$ and $D$.}
\label{fig:multisim}
\end{figure}

The multisimultaneity model was introduced in~\cite{SS97,S97}. The idea is that the frame of interest when a particle undergoes a measurement is its own inertial frame. Within this frame the influences can travel at infinite speed. Thus, when undergoing a measurement, the outcome produced can depend at most on anything that lies in the past of the measurement device. If different measurement stations move relatively to each other, one can arrange a configuration that would not be possible in models with a universal reference frame. In particular, two measurement can be in a before-before configuration if none of them lies in the past of the other one (see Fig.~\ref{fig:multisim}a). The correlations produced in this configuration can only be local.

Here, we also analyse a multisimultaneity model of a second kind by considering, in analogy with $v$-causal models, that particles, at the time they are measured, do not look back at their past in order to produce an outcome, but rather inform their future about the measurement the are undergoing. The situation in which no nonlocal correlation can be produced by two parties then coincides with the situation which was used in~\cite{Scarani02,Scarani05} as an after-after configuration for the original multisimultaneity model (see Fig.~\ref{fig:multisim}b).

Both multisimultaneity models were refuted experimentally using acousto-optic modulators as moving beam-splitters~\cite{ZBGT01,SZGS02,SZGS03,ZBTG01}; an experimental refutation with full measurement stations moving will have to wait for quantum technologies to be implemented on satellites \cite{RJA+12}. Here we show that it can also be refuted using hidden influences inequalities.

In order to show this, consider the space-time configuration depicted in Fig.~\ref{fig:multisim}c or~\ref{fig:multisim}d (depending on the variant of the model chosen). By construction, Bob and Charlie cannot influence themselves in this situation, but they can receive information from Alice and Dave. The condition~\eqref{eq:locality} is thus fulfilled and one is left with checking whether the two points (a') and (b') are satisfied in this configuration.

Seeing that the multisimultaneity model correctly reproduces the Alice-Bob-Dave and Alice-Charlie-Dave marginals here amounts once again to considering that Bob or Charlie could decide to postopone their measurement (see section about finite speed influence models). As for the second condition, one can check directly that any information travelling at the speed of light from $A$ will reach point $D'$ after the information coming at the same speed from $B$, $C$ and $D$, and similarly for $A'$. Both versions of the before-before model thus allow one to communicate faster than light in these configurations which concludes our discussion.

\ \\
\emph{Synthesis. }

The alternative models discussed in this paper aim at describing quantum nonlocality by means of hidden influences propagating or acting within space-time. We proved for all these models that they imply faster-than-light communication and thus are not compatible with relativity. This was possible by applying the inequality presented in~\cite{Bancal12} in relation to the two properties (i) and (ii) which these models satisfy.

Note that the same conclusion could also be reached by using a configurations involving three parties only, with the aid of the recently-discovered tripartite hidden influence inequality~\cite{Tomy}.

It is interesting to observe as well that if one assumes nonlocality at detection~\cite{Suarez12}, then any disappearance of nonlocal coordination between the detectors violates straightforwardly the conservation of energy in the single quantum events. Therefore the alternative models discussed above, as well as those presented in~\cite{Colbeck11}, can be considered refuted by the experiment in~\cite{Guerreiro12}.

\begin{acknowledgements}
We are grateful to Yeong-Cherng Liang and Stefano Pironio for helpful comments and stimulating discussions. This work is supported by the Ministry of Education, the National Research Foundation of Singapore and the Swiss NCCR-QSIT. A. S. acknowledges support from the Social Trends Institute (Barcelona and New York).
\end{acknowledgements}

\end{document}